\documentclass[amssymb,aps,prli,preprint]{revtex4}
\usepackage{latexsym,amsmath,graphics,graphicx,cmbright,exscale,colordvi}
\def\be{\begin{equation}}
\def\ee{\end{equation}}
\def\beq{\begin{eqnarray}}
\def\eeq{\end{eqnarray}}

\begin{document}
\renewcommand{\familydefault}{\sfdefault}
\renewcommand{\sfdefault}{cmbr}

\title{Dynamical magnetic excitations of nanostructures from first-principles}
\author{S. Lounis$^1$}\email{slounis@uci.edu}
\author{A. T. Costa$^{1,2}$}
\author{R. B. Muniz$^2$}
\author{D. L. Mills$^1$}
\affiliation{$^1$ Department of Physics and Astronomy, University of California Irvine, California, 92697 USA}
\affiliation{$^2$ Instituto de F\'isica, Universidade Fedeal Fluminense, 
24210-340 Niter\'oi, Rio de Janeiro, Brazil}

\begin{abstract}
Within time-dependent density functional theory, combined with the Korringa-Kohn-Rostoker Green functions, we devise a real space method to investigate spin dynamics. Our scheme enables one to deduce the Coulomb potential which assures a proper Goldstone mode is present. We illustrate with application to 3$d$ adatoms and dimers on Cu(100).
\end{abstract}
\maketitle
\date{\today}


Very recently new experimental methods for the study of magnetic excitations in the nanoscale and subnanoscale length regime have appeared. Examples are spin polarized electron energy loss spectroscopy (SPEELS)~\cite{zakeri} 
and inelastic scanning tunneling microscopy (STM)~\cite{balashov,heinrich}. 
In both methods, the quantity probed is the transverse susceptibility 
$\chi$ that, in linear response, describes the amplitude of the transverse spin motion produced by an external magnetic field $B_{ext}$ of frequency $\omega$. To calculate $\chi$ is a major computational challenge, if one wants to employ time dependent density functional theory (TD-DFT)~\cite{gross1} or many-body perturbation theory (MBPT) based on DFT\cite{aryasetiawan}. 
Thus one sees very few DFT based calculations even for bulk systems. \cite{savrasov,staunton,sasioglu,buczek}. 
There are studies that use empirical tight binding theory (ETB) \cite{cooke,mills,muniz,costa} based 
on MBPT. In all schemes, a similar master equation must be solved. 
Its solution maybe be written in schematic notation, 
\beq
\chi = \chi_0 (1- U \chi_0)^{-1}
\label{dyson}
\eeq
In ETB MBPT, $\chi_0$ is the single particle susceptibility for a mean field ground state, and $\chi$ is the susceptibility generated by the Random Phase Approximation (RPA). In TD-DFT~\cite{gross1}, $\chi_0$ is the Kohn-Sham susceptibility and 
$\chi$, the total susceptibility, is exact in principle if the full exchange and correlation kernel $U$  
is known. In practice, one often invokes the adiabatic local spin density approximation (ALDA) or rarely one uses  
the RPA approach\cite{sasioglu}; $U$ can also be viewved as a parameter whose value is in the range of $1 eV/\mu_B$ 
for 3$d$ transition elements\cite{himpsel}. 
In the ETB MBPT, $U$ is the effective 
Coulomb interaction that is the local exchange splitting divided by the magnetic moment. Within the ETB where $\chi$ is generated through the RPA 
the Goldstone theorem is satisfied exactly (the scheme is a conserving approximation in many body theory wherein the Ward identity is obeyed) so, for instance, zero wave vector spin waves have precisely zero frequency with spin orbit coupling set aside. In DFT based methods, in practice the Ward identity is not strictly obeyed so the Goldstone theorem is violated in numerical studiess. One then adjusts $U$ in an ad hoc manner so $\chi$ is compatible with the Goldstone theorem. For instance, 
Sasioglu {\it et al.}~\cite{sasioglu} correct $U$ by 45\% in their study of bulk Ni while Buczek {\it et al.} do not calculate it but mention a finite frequency shift for the Goldstone mode in the range of 
$~5-10$ meV~\cite{buczek}. 
This procedure can be dangerous~\cite{future}, and below we introduce a scheme that constrains $U$ so no adjustment are needed.

We report here a new computationally attractive method that allows us to address magnetic excitations without the need to obtain empirical tight binding parameters from other electronic structure calculations. We begin with 
the the Korringa-Kohn-Rostoker single particle Green function (KKR-GF)~\cite{KKR} which contains an ab-initio description of the electronic strucutre, and we introduce an Ansatz that allows us to generate a one electron Green function whose structure is similar to that which enters the ETB methods. We are then led to a computationally tractable and flexible real space scheme for describing spin excitations in nanoscale structures. Along the way we derive a criterion for generating an effective $U$ compatible with the Goldstone theorem. Other approximate schemes have been introduced to address spin excitations within TD-DFT~\cite{buczek} or MBPT based on DFT~\cite{sasioglu}. Within these approaches the calculation of dynamic susceptibilities is still very difficult, and limited to very small systems. Our scheme is readily applied to bulk materials, to surfaces with adsorbed films, and it is a real space formalism ideal for diverse small nanostrucures. As an initial application, below we explore the spin dynamics of single adatoms and dimers on the Cu(100) surface.  For instance, experimentally, Hirjibehedin {\it et al.}~\cite{heinrich} employed the 
semi-insulating CuN substrate that differs qualitatively from the more commonly encountered metal surfaces. Their data can be interpreted with a model Hamiltonian which describes an atomic like localized moment with integer spin, weakly perturbed by the substrate, whereas strong coupling to substrate electronic degrees of freedom qualitatively change the character of the spin excitations in systems we explore here: the magnetic moments are neither integer nor half integer; our approach addresses systems where the model Hamiltonian used in ref.~\cite{heinrich} fails qualitatively.

Our method is based on two assumptions:
(i) if the external applied magnetic field is constant within each atomic cell of it is spherically symmetric in each cell, one can develop a description that maps the calculation onto a scheme discussed many years ago by Lowde and Windsor~\cite{lowde}, though now we have incorporated an ab initio description of the electronic structure into $\chi_0$. This structure, 
as discussed more recently~\cite{muniz}, describes the underlying physics correctly. With our method the size of matrices involved in solving the Dyson equation is modest, thus opening the way to the investigation of large atomic nanosructures with a full description of their electronic structure included. Then (ii) In the KKR-GF method~\cite{KKR}, the Green function is written in terms of products of energy dependent solutions of Schr\"odinger equation in the unit cells; both the regular and irregular solutions enter. We propose here an expansion in terms of energy independent $d$ like wave functions we chose to be the regular solutions of KKR-GF theory evaluated at the Fermi energy. Our focus is on low energy excitations of 3$d$ moments, so this choice is appropriate.

The first step in solving the Dyson equation is to calculate $\chi_{0}$ that is a convolution of two Green functions
\beq
\chi^{ij}_{0}(\vec{r},\vec{r}';\omega)&=&-\frac{1}{\pi}\int dz f(z)
[G^{\downarrow}_{ij}(\vec{r},\vec{r}';z+\omega)
\mathrm{Im}G^{\uparrow}_{ji}(\vec{r}',\vec{r};z)\nonumber\\
&+& \mathrm{Im}G^{\downarrow}_{ij}(\vec{r},\vec{r}';z)
G^{-\uparrow}_{ji}(\vec{r}',\vec{r};z-\omega)]\label{chi0}
\eeq where $f(z)$ is the Fermi distribution function, 
$G$ and $G^-$ represent the retarded and advanced 
one particle Green functions connecting atomic sites $i$ and $j$ and $\mathrm{Im}G=-\frac{i}{2}(G-G^-)$. 
$\chi_0$ can be separated into a sum of two terms: $I_1$ which involves Green functions that 
are analytical in the same half complex plane and a non-analytical $I_2$ \cite{muniz}. For 
positive frequencies:
\beq 
{I}^{ij}_1(\vec{r},\vec{r}';\omega)&=&\frac{i}{2\pi}\int^{E_F} dz [
f(z-\omega)G^{\downarrow}_{ij}(\vec{r},\vec{r}';z)G^{\uparrow}_{ji}(\vec{r}',\vec{r};z-\omega)\nonumber\\
&-&f(z)G^{\downarrow *}_{ji}(\vec{r}',\vec{r};z)G^{\uparrow *}_{ij}(\vec{r},\vec{r}';z-\omega)]
)
\\\mathrm{and} \nonumber\\
{I}^{ij}_2(\vec{r},\vec{r}';\omega)&=&\frac{i}{2\pi}\int_{E_F-\omega}^{E_F} dz 
G^{\downarrow}_{ij}(\vec{r},\vec{r}';z+\omega)
(G^{\uparrow}_{ji}(\vec{r}',\vec{r};z)
-G^{\uparrow *}_{ij}(\vec{r},\vec{r}';z))
\eeq

In order to improve numerical stability, the two terms differ a bit from those 
presented in Ref.\cite{muniz} and they require one less Green function. Such a separation is attractive since 
$I_1$ can be calculated through use of a regular contour in the complex plane~\cite{wildberger} with    
a modest k- and energy-mesh.
$I_2$ can be calculated along a line parallel 
to the real axis that does not require much computational efforts since 
the integration is limited to a small energy regime defined by $\omega$. 
 
The green functions are provided by the KKR-GF method~\cite{KKR}: $G_{ij}(\vec{r},\vec{r}';z)=\sum_{LL_1}-i\sqrt{z}R^{iL}(\vec{r}_<;z)H^{iL}(\vec{r}_>;z)\delta_{ij,LL_1} + 
R^{iL}(\vec{r};z)G^{iL,jL_1}_B(z)R^{jL_1}(\vec{r}';z)$ 
where $G_B$ is the structural Green function. Here  
 the  
regular $R$ and irregular $H$ solutions of the Schr\"odinger equation are energy dependent, and this makes the calculation of $\chi_0$ in Eq.\ref{dyson} tedious and lengthy. Our Ansatz expresses the Green functions in terms of energy independent wave functions $\phi$ such that:
$G_{ij}(\vec{r},\vec{r}';z)\sim \sum_{LL_1}\phi^{iL}(\vec{r})\overline{G}_{ij}^{LL_1}(z)\phi^{jL_1*}(\vec{r}')$ where 
$\overline{G}_{ij}^{LL_1}(z)$ is generated from $\frac{\int\int d\vec{r}d\vec{r}'\phi^{iL*}(\vec{r})G_{ij}(\vec{r},\vec{r}';z)\phi^{jL_1}(\vec{r}')}
{\int d\vec{r}\phi^{iL*}(r)\phi^{iL}(\vec{r}) \int d\vec{r}'\phi^{jL_1}({r}')\phi^{jL_1*}(\vec{r}')}
$. Since the terms in the denominator are normalization 
factors instead of working with $\phi^{iL}(\vec{r})$ we introduce  $\psi^{iL}(\vec{r})={\phi^{iL}(\vec{r})}/
{(\int d\vec{r}\phi^{iL*}(\vec{r})\phi^{iL}(\vec{r}))^{\frac{1}{2}}}$ where 
we choose $\phi(\vec{r})=R^d(\vec{r};E_F)$, {\it i.e.}, 
the $d$-regular solution of the Schr\"odinger equation. This is appropriate for the calculation of the 
$d$-block of the susceptibility.


Within TD-DFT~\cite{gross1}, Eq.~\ref{dyson} has been derived via a 
variational linear response approach. Here, a magnetic system, say magnetized 
initially along the $z$-direction with a magnetization $m_z(\vec{r})$ and 
a charge density $n_0(\vec{r})$, 
is subjected to a small time-dependent 
external transverse magnetic field $B_{ext}(\vec{r};t)$ which induces 
a transverse magnetization $m_{x,y}(\vec{r};t)$. The latter 
quantities are connected to each other through the magnetic response function:
$\chi^{ij}(\vec{r}t,\vec{r}'t')={\delta m^i_{x,y}[B_{ext}](\vec{r}t)/
\delta B^j_{ext}(\vec{r}'t')}|_{B_{ext}=0,n_0}$. Within the atomic sphere approximation (ASA) 
and assuming a spherical external magnetic field this simplifies to 
$\overline{\chi}^{ij}({r}t,{r}'t')=4\pi{\delta m^i_{x,y}[B_{ext}]({r}t)/
\delta B^j_{ext}({r}'t')}|_{B_{ext}=0,n_0}$, where 
$\overline{\chi}^{ij}$ is a sum over all angular momenta components  $\sum_{LL_1}\chi^{iLL_1;jL_1L}$. The same procedure is repeated for 
 the 
magnetic response function $\chi_0$ of the Kohn-Sham noninteracting system 
and after a Fourier transform with respect to time we obtain a form that resembles the 
approach of Lowde and Windsor~\cite{lowde} which is very often used in the tight-binding simulations 
of magnetic excitations~\cite{muniz}. We now have
\beq
\overline{\chi}^{ij}(r,r';\omega)&=&\overline{\chi}^{ij}_{0}(r,r';\omega) \nonumber \\
&+& \sum_{kl} \int dr''\int dr'''
\overline{\chi}^{ik}_{0}(r,r'';\omega) {U^{kl}(r'',r''';\omega)} \overline{\chi}^{lj}(r''',r;\omega)
\eeq 
that involves only site dependent matrices. $U^{ij}(r,r';\omega)$ is a functional derivative given by 
$\frac{\delta B^i_{eff}(r;\omega)}{4\pi\delta m^j(r';\omega)}|_{B_{ext}=0,n_0}$ 
that simplifies within ALDA~\cite{katsnelson} to 
$\frac{B^i_{eff}(r; 0)}{4\pi m^i_z(r;  0)}\delta_{r,r'}\delta_{i,j}$ assuming an initial collinear magnetic configuration; $B_{eff}$ is the
 magnetic part of the effective Kohn-Sham potential ($V_{eff}^{\downarrow}-V_{eff}^{\uparrow}$). $U$ can be considered as 
an exchange splitting divided by the magnetization. 

It is interesting to note that there is another way to determine the 
correct form of $U$. Indeed, to ensure the realization of the Goldstone mode, 
it is necessary to have the following sum rule fulfilled at $\omega = 0$~\cite{future}:
\beq
\sum_j\int d\vec{r}' \chi^{ij}_0(\vec{r},\vec{r}';0)B^j_{eff}(\vec{r}';0)
&=& m^i_z(\vec{r};0)
\label{sumrule}
\eeq 
this expression has been obtained after multiplying both sides of Eq.~\ref{chi0} by $B_{eff}^j(\vec{r}';\omega=0)$, integrating over 
$\vec{r}'$, summing up over all sites j and using the following Dyson equation:  $G_{\uparrow}=G_{\downarrow}+G_{\downarrow}B_{eff} G_{\uparrow}$. Eq.~\ref{sumrule} can be rewritten in a matrix notation:
\beq
U^i=\Gamma^{-1}m^i_z
\label{sumrule2}
\eeq
with $\Gamma=\chi_0m_z$ at $\omega=0$ and $U^i({r})=\frac{B_{eff}({r};0)}{4\pi m^i_z({r};0)}$. We then recover the 
result extracted from the ALDA: Eq.~\ref{sumrule2} provides a means of calculating a value of $U$ which leads to satisfaction of the Goldstone theorem. The right hand side requires knowledge of the magnetization and 
the static susceptibility. Stated otherwise, the correct $U$ is the one 
with the lowest eigenvalue of the denominator of Eq.\ref{dyson} associated with the magnetic moments as components of the eigenvectors. 
 Such a derivation is advantageous since it would work 
for more complex systems involving non-equivalent atoms.

Assuming the expansion in terms of energy independent wave functions described previously, the final Dyson equation simplifies after some straightforward 
algebra into a strictly site dependent equation  
\beq
\overline{\overline{\chi}}=\overline{\overline{\chi}}_0 + 
\overline{\overline{\chi}}_0U\overline{\overline{\chi}}
\eeq
where the d-block of the dynamical susceptibility $\overline{\chi}_0(r,r';\omega)$ 
is given by $\psi^{id}_{\downarrow}(r)\psi^{id*}_{\uparrow}(r)
\overline{\overline{\chi}}_0(\omega)\psi^{jd*}_{\downarrow}(r')\psi^{jd}_{\uparrow}(r')$ and U can be calculated once for every atom from the previous sum rule, 
Eq.~\ref{sumrule2}.
 It can be understood as a Stoner parameter 
and gives once more a justification for the 
approach used by Lowde and Windsor\cite{lowde}: i.e. the effective intra-atomic Coulomb interaction is expressed 
by only one parameter.

 To illustrate our scheme, 
we shall investigate the magnetic excitations of 3$d$ adatoms and  
dimers deposited on Cu(001) surface 
with the theoretical LDA lattice parameter (6.64 a.u.). First we examine the 
spin dynamics of the magnetic moment bearing adsorbates Cr, Mn, Fe and Co.
The atoms are positioned in the fourfold hollow sites with the 
first nearest neighbors around every adatom included 
when evaluating the full real space Green function.
\begin{table}[ht!]
\begin{center}
\caption{\label{comparison}Comparison between magnetic moments (in $\mu_B$) of adatoms calculated from the KKR-GF method or following the projection scheme discussed in the text. Values of $U$'s (eV/$\mu_B$) are also shown.}
\begin{ruledtabular}
\begin{tabular}{lcccc}
                                   & Cr       & Mn           &  Fe         & Co\\
\hline
 $M_d$: KKR-GF/model                      &4.0/4.0      &  4.1/3.9        &  3.00/2.7  & 1.7/1.6\\
 $M_{total}$                        &4.0      &  4.2        &  3.0       & 1.7\\
 -$U$             &0.90      &  0.97        &  0.97       & 0.97%
\end{tabular}
\end{ruledtabular}
\end{center}
\end{table}
 
\begin{figure}
\begin{center}
\includegraphics*[angle=0,width=1.\linewidth]{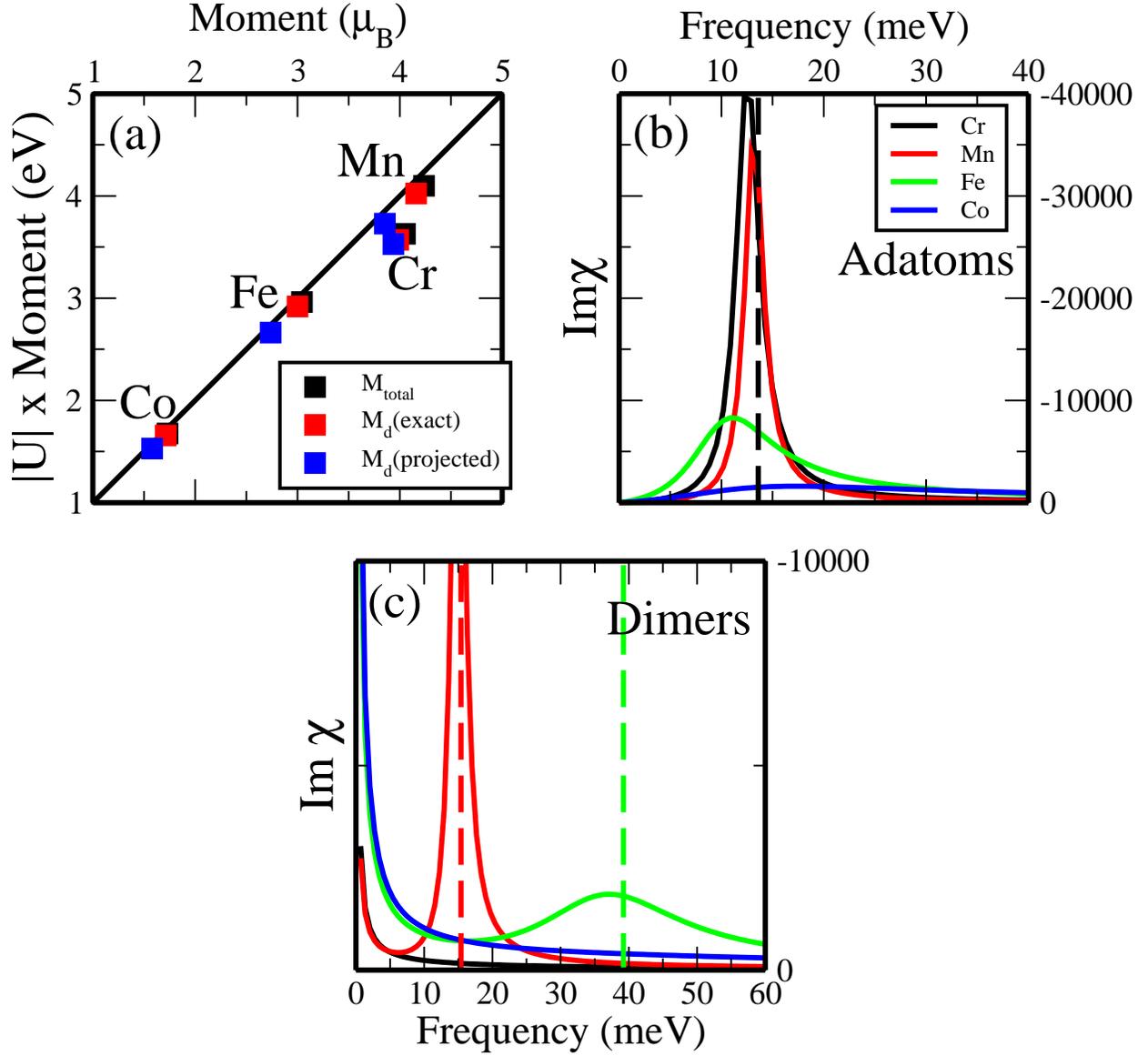}
\end{center}
\caption{In (a) is shown the variation of the product of $U$ with the different 
values of the moments. The values obtained for the adatoms fall on the diagonal as estimated in 
the curve of Himpsel\cite{himpsel}. In (b) the imaginary part of 
$\chi$ is plotted for every adatom when applying an additional magnetic field along the $z$-direction corresponding to a Larmor frequency of 13.6 meV (dashed line).
In (c) are shown Im$\chi$ calculated for the four dimers. The optical 
modes, estimated for Mn and Fe from a Heisenberg model, are represented as dashed lines.}
\label{compilation}
\end{figure}

As shown in Table \ref{comparison}, the $d$-contribution to the total moment is the most relevant and 
is nicely reproduced by the projection of the Green functions into our 
choice of wave functions. The $U$'s calculated are very close to 1eV/$\mu_B$, a 
value that is  
very often used in the literature~\cite{muniz} which was extracted empirically from data by Himpsel~\cite{himpsel}. Himpsel related the energy splitting between majority and minority spin states observed in photoemission to the magnetic moment of 3d atoms.  In Fig.~\ref{compilation}(a), we plot the energy splitting between our calculated majority and minority spin states as a function of magnetic moment, to find a linear relation with slope very close to that in the corresponding figure in Himpsel's paper~\cite{himpsel}. This 
provides us with confidence in the scheme set forth in this paper.  

We note that many years ago the spin dynamics of a moment embedded in a paramagnetic host was described~\cite{lederer} within a framework conceptually similar to that used here. The local response displays a g shifted Zeeman resonance, broadened very substantially by decay of the coherent spin precession to particle hole pairs, whereas the total moment of the system precesses with g=2 and zero linewidth. Our calculations explore the local response of the moment. Local probes such as STM thus examine aspects of spin dynamics inaccessible to spectroscopy based on excitation by long wavelength radiation such as microwaves or light; these sense the precession of the total moment.

In Fig.~\ref{compilation}(b), we show our calculations of the resonant response of the local moments for the four adatoms we have investigated. The resonant frequency scales linearly with the applied DC field, as does the width of the structure. For this reason, as in an earlier study~\cite{muniz}, 
we can apply a DC field of sufficient strength to move the resonance up to frequencies that allow numerical study. A simple scaling reduces the results to the regime of physical applied fields. The g shift can be positive (Co) or negative (Cr, Mn and Fe). (For the field we have applied, a g value of 2 would provide a resonance at 13.6 meV.) These results agree well with earlier tight-binding simulations performed on Mn and Co adatom ~\cite{muniz}. 
The width of the resonances is controlled by the local density of states~\cite{lederer}, and is thus influenced by the position of the d levels relative to the Fermi energy. Thus the Co and Fe resonances are quite broad, since their minority spin levels intersect the Fermi level, whereas those for Mn and Cr are much sharper since for these adatoms the Fermi level lies between the majority and minority states.

We have also explored nearest neighbor dimers of the 3d adatoms discussed above, assuming a ferromagnetic ground state. For the dimer, there are two resonances, a zero frequency acoustical mode, and a high frequency optical mode that is damped by decay to Stoner excitations. The position of the optical mode provides information on the stability of the assumed ground state. We find the optical mode at negative frequency for Cr and Co. This informs us the ferromagnetic ground state is unstable, whereas for the Fe and Mn dimer the mode resides at positive frequency so ferromagnetism  is stable.  This test of ground state stability will be useful for studies of more complex structures. We have calculated adiabatic exchange integrals $J$ in an effective Hamiltionian we write as $-J\vec{e}_1\cdot \vec{e}_{2}$, 
with $\vec{e}_1$ and $\vec{e}_2$ unit vectors.~\cite{LKAG} We find $J$ 
negative for Cr ($J=-19.8$ meV) and Co ($J=-14.9$ meV), consistent with instability of ferromagnetism in the ground state, while positive 
values were obtained for Fe ($J=30.4$ meV) and Mn ($J=16.3$ meV) so we 
confirm ferromagnetism is stable. For positive frequencies  we plot the imaginary part of the susceptibility in Fig.~\ref{compilation}(c). For Cr, Mn, Fe and Co the Heisenberg model gives -19.7 meV, 15.4 meV, 39.2 meV and -33.1 meV for the optical mode frequencies respectively.

To conclude, we have shown that a simple approach, based on TD-DFT and the KKR-GF method, can be used to extract dynamics magnetic susceptibilities. This eliminates the need to acquire the parameters which enter empirical tight binding approaches, since a description of the electronic structure is embedded into our single particle Green functions. The size of matrices  in the Dyson equation are small enough to permit calculations of large nanostructures in future applications. We have developed an identity that leads to a numerically stable method of  extracting the proper Coulomb interaction to be used in the Dyson equation from which the full dynamical susceptibility is obtained. As an application, 3$d$ adatoms and dimers deposited on Cu(001) surface were investigated from 
first-principles.

Research supported by the U. S. Depart. of Energy through grant No.
DE-FG03-84ER-45083. S. L. thanks the A.v. Humboldt Foundation 
for a Feodor Lynen Fellowship. R.B.M. acknowledges support from CNPq and
FAPERJ, Brazil.

\end{document}